\documentclass{emulateapj}

\slugcomment{}

\shorttitle{The Expanded Very Large Array}
\shortauthors{Perley, Chandler Butler \& Wrobel}

\begin{document}

\title{The Expanded Very Large Array -- a New Telescope for New Science}

\author{R. A. Perley\altaffilmark{1}, C.J.Chandler\altaffilmark{1},
  B.J.Butler\altaffilmark{1}, and J.M. Wrobel\altaffilmark{1}}

\altaffiltext{1}{National Radio Astronomy Observatory, P.O. Box O,
Socorro, NM 87801; rperley@nrao.edu, bbutler@nrao.edu,
cchandle@nrao.edu, jwrobel@nrao.edu}

\begin{abstract}
Since its commissioning in 1980, the Very Large Array (VLA) has
consistently demonstrated its scientific productivity.  However, its
fundamental capabilities have changed little since 1980, particularly
in the key areas of sensitivity, frequency coverage, and velocity
resolution.  These limitations have been addressed by a major upgrade
of the array, which began in 2001 and will be completed at the end of
2012.  When completed, the Expanded VLA -- the EVLA -- will provide
complete frequency coverage from 1 to 50 GHz, a continuum sensitivity
of typically 1 $\mu$Jy/beam (in 9 hours with full bandwidth), and a
modern correlator with vastly greater capabilities and flexibility
than the VLA's.  In this paper we describe the goals of the EVLA
project, its current status, and the anticipated expansion of
capabilities over the next few years.  User access to the array
through the OSRO and RSRO programs is described.  The following
papers in this special issue, derived from observations in its early
science period, demonstrate the astonishing breadth of this most
flexible and powerful general-purpose telescope.
\end{abstract}


\section{Introduction}\label{introduction}

\begin{figure}[b!]
\centerline{\hbox{
\includegraphics[width=8.5cm]{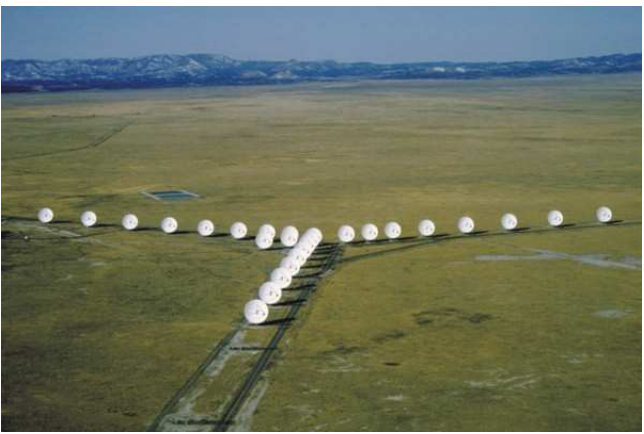}}}
\centerline{\parbox{8.5cm}{
\caption{\small An aerial photograph of the VLA in its most compact {\bf ``D''}
  configuration, with maximum baseline 1.0 km. The array can be
  reconfigured by moving the antennas to more distant pads via the
  rail lines visible in the photograph.}
\label{fig:VLAPic}}}
\end{figure}

The Very Large Array (VLA) is an imaging radio interferometer located
in west-central New Mexico, operated by the National Radio Astronomy
Observatory (NRAO)\footnote{The National Radio Astronomy Observatory
is a facility of the National Science Foundation operated under
cooperative agreement by Associated Universities, Inc.}.  It comprises
27 antennas of 25-meter diameter positioned along three equiangular
arms of length 21 km, nine antennas per arm.  The array provides
images of astronomical objects in all four Stokes parameters, with a
diffraction-limited maximum resolution of 1.4 arcseconds at 1.4 GHz
and 40 milliarcseconds at 50 GHz, utilizing the well-established
techniques of aperture synthesis, as described for example in
\cite{tho01}.  The VLA is an exceptionally flexible telescope, in part
due to its ability to reconfigure -- there are four standard
configurations of maximum baseline lengths of 1, 3.4, 11, and 36 km,
providing a wide range of resolutions and image surface brightness
sensitivities.  Descriptions of the VLA as originally designed are
given found in \cite{tho80} and \cite{nap83}.  A picture of the VLA
in its most compact configuration is shown in Fig.~\ref{fig:VLAPic}

The VLA was designed and built in the 1970s utilizing the best
technology available at the time.  Upon its completion in 1980, the
array could observe in four frequency bands with a maximum bandwidth
of 100 MHz per polarization.  Its innovative digital correlator
provided a maximum of 512 spectral channels, spanning a maximum of 3
MHz for each of the 351 baselines in a single polarization, or it
could provide full Stokes visibilities for polarimetric imaging but
without any spectral resolution.  These capabilities were
ground-breaking at the time, and were well-matched to the key science
goals for the telescope, which included imaging the Doppler-shifted
hydrogen emission from nearby galaxies, and resolving the fine-scale
structure of powerful radio galaxies, quasars, and supernova remnants.

With 27 antennas and a 2-dimensional array, the VLA was designed for
sensitivity, speed, and flexibility of operation.  Besides the ability
to change its physical scale, changes in frequency band and
corrrelator mode could be effected in seconds, enabling astronomers to
acquire a range of information on astronomical objects not possible on
any other centimeter-wavelength radio telescope.  These attributes
encouraged users of the VLA to image radio emission from processes far
removed from those given in the original proposal as science goals for
the array -- indeed, much of the VLA's most original and influential
observations are in fields unanticipated or unknown at the time of
construction.

Most components of the VLA's design remained unchanged for twenty
years following its completion -- in particular, the signal
transmission and correlation capabilities remained at their 1980
levels, essentially freezing the array's bandwidth and spectral
resolution.  During this interval, and continuing on to this day,
spectacular improvements in signal transmission and processing have
taken place, making it clear that a minimum of an order of magnitude
improvement in the array's sensitivity, frequency coverage and
spectral resolution could be obtained at modest cost by implementing
these new technologies.  During this same interval, the breadth and
range of astronomical science had changed dramatically, with ever
increasing emphasis on rapid time response, fast imaging, precision
polarimetry, high sensitivity, wider frequency coverage, and higher
spectral resolution.  In response to these expanding needs, the NRAO
proposed to the NSF in 2000 a far-reaching expansion of the VLA's
capabilities, essentially marrying modern high-speed wide-band digital
and wide-band receiver technologies to the sound infrastructure
already in place.  Following a high ranking by the 2000 Decadal Review
\cite{astro2001}, the EVLA Project began in 2001 and will be completed
by the end of 2012.  The Expanded Very Large Array Project is a
partnership among the U.S., Canada, and Mexico, with a total budget
of \$96M (inflation-adjusted, in 2011 dollars).

\section{Key EVLA Goals and Capabilities}\label{EVLA Capabilities}

The technical goals of the EVLA are based on a comprehensive review of
potential science enabled by a minimum tenfold increase in
capabilities over the VLA\@.  The identified science capabilities were
organized into four major themes:
\begin{itemize}
\item The magnetic universe:  measuring the strength and topology of
  cosmic magnetic fields;
\item The obscured universe: enabling unbiased surveys and imaging of
  dust-shrouded objects that are obscured at other wavelengths;
\item The transient unverse: enabling rapid response to, and imaging
  of, rapidly evolving transient sources;
\item The evolving universe: tracking the formation and evolution of
  objects in the universe, ranging from stars through nearby galaxies
  to galactic nuclei.
\end{itemize}

Within each theme, it was readily demonstrated that order-of-magnitude
improvements in VLA performance would result in spectacular new
science by the world-wide user community.  Based on these conclusions,
the fundamental goals of the EVLA Project are:
\begin{itemize}
\item Complete frequency coverage from 1 to 50 GHz, via eight new or
  improved receiver bands, utilizing state-of-the-art technology. See
  Table~\ref{tab:EVLASens} for basic characteristics. 
\item New antenna electronics to process eight signal channels of up
  to 2 GHz each.
\item High-speed wide-band digital samplers within each antenna to
  sample up to 8 GHz bandwidth in each polarization.  
\item A new wide-bandwidth fiber-optical data transmission system to
  conduct digital signals from the antennas to the correlator.  
\item A new wide-bandwidth, full polarization correlator providing a
  minimum of 16384 spectral channels per baseline.     
\item A new real-time control system for the array, and new monitor
  and control hardware for the electronics.
\item New high-level software that provides easier and more flexible
  management of EVLA capabilities to its users.
\end{itemize}

\begin{deluxetable}{cccccc}
\tabletypesize{\scriptsize} \tablecolumns{6} \tablewidth{0pc}
\tablecaption{EVLA Band Characteristics\label{tab:EVLASens}}
\tablehead{ \colhead{} & \colhead{Letter} & \colhead{Available} &
\colhead{Antenna} & \multicolumn{2}{c}{Sensitivity\tablenotemark{a}}\\
\colhead{Band} & \colhead{Code} & \colhead{Bandwidth\tablenotemark{b}}
& \colhead{SEFD\tablenotemark{c}} & \colhead{Continuum} &
\colhead{Line} \\ \colhead{(GHz)} & \colhead{} & \colhead{(GHz)} &
\colhead{(Jy)} & \colhead{($\mu$Jy/bm)} & \colhead{(mJy/bm)} } \startdata
1--2 & L & 0.7 & 400 & 5.5 & 2.2\\ 2--4 & S & 1.75 & 350 & 3.9 & 1.7\\
4--8 & C & 3.5 & 300 & 2.4 & 1.0\\ 8--12 & X & 3.8 & 250 & 1.8 &
0.65\\ 12--18 & Ku & 5.5 & 280 & 1.7 & 0.61\\ 18--26.5& K & 8 & 450 &
2.3 & 0.77\\ 26.5--40& Ka & 8 & 620 & 3.2 & 0.90\\ 40--50 & Q & 8
&1100 & 5.6 & 1.4\\ \enddata \tablenotetext{c}{The System Equivalent
Flux Density is a measure of the antenna sensitivity: ${\rm
SEFD}=2kT_{sys}/A_e.$ It is the flux density of a source which doubles
the system temperature.}  \tablenotetext{b}{An estimate of the
effective bandwidth available, free of RFI} \tablenotetext{a}{The
expected rms noise in a 1-hour integration at high elevation and under
good weather conditions.  For the Continuum case, the bandwidth
utilized is that listed in column four.  For the Line case, a
bandwidth corresponding to 1 km/sec velocity resolution is assumed.}
\end{deluxetable}

\subsection{The WIDAR Correlator}

A key component of the EVLA is its new wideband digital correlator,
known by the acronym WIDARS\footnote{For Wideband Interferometric
Digital ARchitecture}.  This is a 10 peta-32-bit ops/sec
special-purpose computer which produces the cross-power spectral
visibilities for all baselines in the array.  A description of
its design is given in \cite{car00}.  Its key astronomical
capabilities are summarized below:
\begin{itemize}
\item 16 GHz maximum instantaneous input bandwidth
\item A minimum of 16384 spectral channels per baseline, and a maximum
  exceeding 4 million.  
\item Full polarization capabilities on all baselines and channels. If
  full polarization is not needed for the science, correlator resources
  can be reallocated to provide higher spectral resolution for the
  parallel-hand correlations.
\item Generation of 64 independent ``spectral windows''\footnote{A
  spectral window is a digitally defined frequency span, which is
  uniformly subdivided into spectral channels by the correlator.},
  each of which is separately tunable in frequency and bandwidth.  The
  spectral window width is variable, by factors of two, from 128 MHz
  down to 31 kHz.
\item The ability to improve spectral resolution by utilizing
  correlator resources freed up with a reduction of the spectral
  window width, or by reallocating resources from unneeded spectral
  windows or polarization products.  The spectral resolution is
  adjustable from a maximum of 2000 kHz to a minimum of 0.12 Hz.
\item A minimum integration time of 100 msec with the standard minimum
  16384 spectral channels, and less with a reduced spectral
  resolution.
\end{itemize}
Besides these basic capabilities for regular observing modes, there
are a number of speciality modes, including:
\begin{itemize}
\item A phased array mode, where the signals from all antennas are
  combined in phase, and made available for external capture and analysis,
  such as for VLBI applications and pulsar processing.
\item A specialized pulsar binning mode, providing up to 2000 phase
  bins with temporal resolution as short as 200 $\mu$sec with all
  spectral channels, and as short at 15$\mu$sec with a reduced
  spectral resolution, enabling rapid imaging of objects such as
  globular clusters and the Galactic Center where multiple pulsars are
  expected to lie within the antenna primary beam.
\item Up to eight simultaneously-operating subarrays, each with a
  different target point and correlator configuration.  
\item An external data capture capability, allowing antenna or phased
  array outputs to be externally recorded for off-line processing.
\end{itemize}

The WIDAR correlator is Canada's contribution to the EVLA project, and
was designed and built to meet or exceed the requirements of the EVLA
by the HIA correlator group, located at DRAO near Penticton, BC,
Canada.

A more thorough description of the EVLA's design, including that of
its correlator, is found in \cite{per09}.
\section{EVLA Capabilities Growth}\label{capabilities growth}

A compact summary of the expansion in capabilities of the EVLA in
comparison to those of the VLA is provided in Table~\ref{tab:EVLA}

\begin{deluxetable}{cccc}
\tabletypesize{\scriptsize}
\tablecolumns{4}
\tablewidth{0pc}
\tablecaption{EVLA-VLA Performance Comparison\label{tab:EVLA}}
\tablehead{
\colhead{Parameter} &
\colhead{VLA}       &
\colhead{EVLA}      &
\colhead{Ratio}     }
\startdata
Continuum Sensitivity&10$\mu$Jy//bm&1$\mu$Jy/bm&10\\
Maximum BW, per polarization&0.1 GHz&8 GHz&80\\
\# of freq. channels at max. BW&16&16384&1024\\
Max. \# of freq. channels&512&4194304&8092\\
Coarsest frequency resolution&50 MHz&2 MHz&25\\
Finest frequency resolution&381 Hz&0.12 Hz&3180\\
\# of full-pol. spectral windows&2&64&32\\
Frac. Freq. coverage (Log scale)&22\%&100\%&5\\
\enddata
\end{deluxetable}

The conversion of the VLA into the EVLA was scheduled to take more
than a decade.  It was therefore considered vital to maintain
operation as a productive scientific facility throughout the
conversion process.  This required designing in a backward
compatibility between the newly-converted antennas and the original
correlator.  This process has been very successful, enabling nearly
seamless continuing observing, with the array only shut down for a
single 7-week period between January and March 2010 in order to move
hardware from the old VLA correlator to the WIDAR correlator upon the
latter's implementation.  This has enabled the NRAO to offer steadily
increasing scientific capabilities to the user community ahead of the
completion of the construction project.  The growth in capabilities
can be separated into two parts: that provided by the antennas,
including the receivers and the data transmission system, and that
provided by the correlator.
  
\subsection{Antenna and Frequency Band Capabilities}\label{receivers}

\begin{figure}[h]
\centerline{\hbox{
\includegraphics[width=8.5cm]{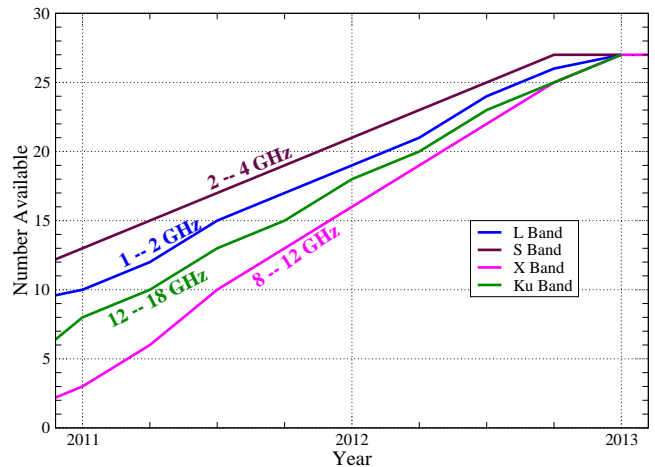}}}
\centerline{\parbox{8.5cm}{
\caption{\small The predicted availability of wideband EVLA receivers
  for the remaining four receiver bands.  The 6--8 GHz band, and the
  three bands spanning 18 through 50 GHz are now fully outfitted.}
\label{fig:RcvrAvail}}}
\end{figure}

Figure~\ref{fig:RcvrAvail} shows the current and anticipated
availability of the eight receiver bands.  Full outfitting of four
receiver bands are now complete -- these are the 4 -- 8 GHz band, and
the three highest frequency bands, spanning 18 through 50 GHz.  

A critical component, not illustrated in the figure, is the growth in
data transmission capabilities.  The current maximum total bandwidth
which can be transferred from antenna to correlator is 4 GHz,
available in two pairs of oppositely polarized signals of bandwidth 1
GHz each.  Implementation of the full 16 GHz capability will not be
available for science observing until at least mid-2012.  

\subsection{Growth in Correlator Capabilities}

The WIDAR correlator is capable of a diverse range of observing modes.
Following the decommissioning of the old VLA correlator in January 2010,
a basic set of WIDAR correlator capabilies were defined and offered
for the first full EVLA standard configuration cycle that were
modelled on the capabilities of the VLA, and in fact more than doubled
the total available bandwidth per polarization, to 256 MHz, and
dramatically increased the number of available spectral channels at
the maximum bandwidth.  At the same time the commissioning of the
correlator proceeded to focus on delivering the maximum 2 GHz per
polarization bandwidth currently available in preparation for the
configuration cycle beginning in September 2011 in the most compact
configuration.  The jump from 256 MHz to 2 GHz marks a potential
increase in dataset size by a factor of 8, and with the total 8 GHz
bandwidth expected in 2012, there will be a further increase.  In its
highest spectral resolution modes, with typically hundreds of
thousands of channels per integration, visibility dataset sizes will
potentially be several terabytes.  By the beginning of EVLA full
operations in January 2013, it is expected that up to 8 GHz bandwidth
(for those bands supporting these bandwidths; see Table 1) will be
available, with up to 4 million channels for spectroscopy.

\subsection{Science Commissioning}

The delivery of the full science capabilities of the EVLA, as opposed
to the correlator modes and updated electronics and receivers,
requires the development of new observing, calibration, and
post-processing procedures compared with the VLA, plus the software to
support them.  The WIDAR correlator is fundamentally a spectral line
correlator, so that even standard ``continuum'' observations of
astronomical sources are carried out in a spectral line correlator
mode.  Furthermore, the vastly increased sensitivity of the EVLA
provides new opportunities such as fast ``on-the-fly'' mosaics of
large areas, large (in number) source surveys, and high time
resolution observations.  All these observing modes are new to the
EVLA, and each will be commissioned by EVLA staff.  At the time of
EVLA full operations many new observing modes and capabilities are
expected to be available, but some more specialized modes may take
longer.

The wide bandwidths of the EVLA present a special problem at its lower
frequencies ($\nu \la 10$~GHz, although higher frequencies are also
affected to some extent), for which the radio spectrum suffers
considerable external interference (RFI) that is both temporally and
spatially variable.  The development of automated flagging and
interference excision procedures and software is a key area of
commissioning throughout 2011 and 2012.

\subsection{EVLA Early Science Programs}

EVLA Early Science began in March 2010 with the array in its most
compact, D-configuration, and will continue through the end of the
EVLA construction project until full operations commences in January
2013.  It includes an Open Shared Risk Observing (OSRO) program for
the general user community, and a Resident Shared Risk Observing
(RSRO) program.  The OSRO program offers the correlator capabilities
described in Section 3.2, above.  The RSRO program offers participants
full access to the growing capabilities of the WIDAR correlator for
peer-reviewed science projects, in exchange for a period of residence
at the Domenici Science Operations Center (DSOC) in Socorro to assist
with the EVLA commissioning.  It is intended to accelerate the
development of the EVLA's full scientific capabilities by gaining
enhanced resources and expertise through community participation, at
the same time as more quickly optimizing the scientific productivity
of the EVLA\@.  To date, 27 individuals have passed through the DSOC
as RSRO visitors.  The papers in this Special Issue of the
Astrophysical Journal Letters utilize data obtained through both of
these Early Science programs.

\section{Using the EVLA}\label{EVLA Usage}

Observing time on the EVLA is open to all astronomers world-wide.
There are no quotas or reserved blocks of time. Starting in 2011 time
on the EVLA is scheduled on a semester basis, with each semester
lasting six months. Proposal deadlines are 5pm (1700) Eastern Time on
February 1 and August 1.\footnote{It is also possible to obtain EVLA
observing time by proposing to NASA missions, under joint agreements
between NRAO and those missions. Such programs currently exist for the
Chandra and Fermi missions; consult the relevant mission proposal
calls for more information.} The February 1 deadline nominally covers
observing time in August or later, and the August 1 deadline nominally
covers observing time in February or later. At either proposal
deadline, requests for future array configurations may also be
considered.

Astronomers prepare and submit observing proposals using the on-line
Proposal Submission Tool (PST). The PST permits the detailed
construction of a cover sheet specifying the requested observations,
using a set of on-line forms, and the uploading of a scientific and
technical justification to accompany the cover information. Funding
opportunities are available for students at US institutions and may be
requested via the PST. Proposal evaluation involves technical reviews
by NRAO staff and panel-based science reviews by community
members. The results of these reviews are cross-reconciled by the
community-based Time Allocation Committee, leading to an approved
science program.

Information from approved proposals flows to the web-based Observation
Preparation Tool (OPT), an on-line tool for creating observing
scripts. Astronomers use the OPT to specify sources to be observed,
instrumental setups, and timing information, all packaged as a
Scheduling Block (SB). Astronomers also use the OPT to submit their
Scheduling Blocks to a dynamic queue. NRAO staff use the Observation
Scheduling Tool (OST) to examine the queue of pending scheduling
blocks and the current observing conditions. The OST then applies
heuristics to select the optimal scheduling block and send it off for
execution by the monitor and control system. Such dynamic scheduling
enhances science data quality and the array's ability to discharge
time-sensitive science. The OST can also be used to execute a
fixed-date scheduling block, as required for a coordinated observation
with other telescopes.

As an observation progresses, NRAO staff monitor the array's health,
maintain an observing log, and ensure that science data are being
archived. At the conclusion of an observation the observing log is
e-mailed to the astronomer. This log includes a link to the Archive
Access Tool (AAT), an on-line search tool that the astronomer can use to
locate and download the archived data. Those data are proprietary to
the proposal's authors for a period of 12 months.

\section{Data Post-Processing, Pipelines, and Algorithm Development}

It is clear that the data post-processing needs of the EVLA will far
exceed those of the VLA.  The salient features of the EVLA that will
drive the post-processing software development are the capabilities to
produce data covering: (i) a large instantaneous bandwidth (at the
lowest three frequency bands the high-frequency end is twice the
frequency of the low end), and (ii) a large number of spectral
channels (usually flexibly arranged in non-contiguous multiple
spectral windows of varying width and frequency resolution).  These
two features result in high data rates ($>$ 5 MB/s) as a matter of
course, with the WIDAR correlelator capable of producing much higher
rates (up to 350 GB/s!), and also therefore with the prospect of
dealing with large ($>$ 1 TB) datasets.  Therefore, calibration and
imaging of the data from the EVLA will present problems that must be
solved by a combination of a scalable post-processing package,
algorithmic improvements, and processing and I/O speed gains.

A large fraction of data collected by the EVLA will be taken in one of
a number of standard observing modes, for example, low frequency
continuum, high frequency continuum, HI (neutral hydrogen) spectral
line, or polarization.  Given the considerable experience in reducing
data taken in similar kinds of modes with the VLA, it is reasonable to
assume that reduction of this type of data can be mostly automated.
The post-processing package (see below), when combined with some
information collected from the astronomer in the observation
preparation stage (in the Observation Preparation Tool - for instance
what a particular source is meant to actually be used for in the
post-processing), during actual observing, and with some heuristics
(rules for what to do given certain situations) should be sufficient
to complete such automatic reductions.  While we are not currently
providing an automated reduction pipeline for EVLA data, we plan to do
so in the near future.  When this occurs, all data taken in any of the
standard modes on the EVLA will be processed with this pipeline, and
the results (mainly so-called ``reference image cubes'') made
available via the science data archive, subject to proprietary
constraints.  While these reference image cubes may be sufficient to
give the investigators (and others) an idea of data quality and crude
source characteristics, there is some concern that it will be
extremely difficult to ever provide completely reliable automatic
pipeline products as a {\it final} data product from the EVLA.  For
that, some human intervention, either by NRAO staff, or the
astronomers themselves, may be needed.  This is an active area of
investigation within the observatory.

For all data which cannot be reliably reduced via a pipeline, or for
astronomers who wish to modify or extend what is done within the
pipeline, there must be a post-processing software package capable of
performing all steps necessary to turn the measured visibilities into
final image cubes.  For the VLA, several packages have been used for
data editing and calibration over the years, but for nearly the entire
lifetime of the VLA, AIPS (\cite{grei03}) has been the primary software
package for data processing.  There are problems with AIPS, however,
that prevent its continued long-term use for EVLA data
post-processing.  The post-processing package of choice for the EVLA
project is CASA (Common Astronomical Software Applications, see, e.g.,
\cite{McMul07}), because of its scalability, ability to be parallelized,
scriptability, expertise within NRAO, and commonality with ALMA.  Not
everything that is needed for EVLA data reduction (and certainly not
for robust pipeline-reduction) is currently available within CASA,
however, so there is a program of implementation of missing pieces
within the package.  In addition, a vigorous program of algorithm
development within NRAO is ongoing, to address items that are not only
not implemented within CASA, but have no generally accepted
algorithmic solution at all.  Notably, automatic flagging of suspect
data, wide-field wide-bandwidth full polarization imaging, RFI
detection and excision, and ionospheric corrections are all areas of
active research, and will be implemented within CASA as soon as
possible after accepted algorithms are developed.  Many of these areas
of research are of course not unique to the EVLA, so developments at
other observatories and telescopes are closely watched so that we can
take advantage when appropriate.

Finally, in order to support the scale of computing that is needed for
both pipeline and hands-on post-processing of EVLA data, we are
planning to provide a mid-sized computing cluster (10's of nodes) for
both the automatic pipeline reduction of standard mode observations,
and for somewhat more interactive reduction by astronomers.  We
believe this, along with improvements in the speed of the CASA code
itself, is sufficient to support the needs of the astronomical
community.  We assume that most access to the cluster will be either
by the NRAO-controlled automatic reduction pipeline, or by batch jobs
submitted by remote users.  We are committed to remaining flexible in
this plan, however, as this will be a new era for post-processing of
interferometer data, and we must be able to react to new developments,
pressure from the community, and other realities as we go forward.

\section{Summary}

The EVLA is a major expansion to the highly flexible and productive
VLA.  By expanding the bandwidth to 8 GHz/polarization, adding
receivers to provide full frequency coverage from 1 to 50 GHz, and
implementing a new correlator with superb spectral resolution and
flexibility, the EVLA will provide orders of magnitude improvement in
scientific capability over the VLA -- capabilities that will ensure
that the EVLA will be the premier general-purpose cm-wave imaging
radio telescope for at least the next decade, serving the world user
community for investigations into, and understanding of, celestial
radio transients, the evolution of radio-emitting objects in the
universe, the structure and strength of celestial magnetic fields, and
the dusty obscured regions opaque to most other wavelengths.  

\acknowledgments A project of this magnitude requires the talents and
dedication of hundreds of individuals, far too numerous to list here.
To all of these we offer our thanks for their efforts to make this
project a success.  It is a pleasure to acknowledge the three funding
agencies supporting this project: The U.S. National Science
Foundation, the Canadian National Research Council, and the Mexican
Consejo Nacional De Cencia y Technologia.

{\it Facilities:} \facility{VLA}


\begin{thebibliography}{Thompson, A.R., Clark,
 B.G., Wade, C.M., and Napier, P.J. {\it Astrophys. J. Suppl.}, 44,
 151--167, 1980.}
\bibitem[Thompson, Moran, and Swenson (2001)]{tho01} Thompson, A.R.,
  Moran, J.M., and Swenson, G.W., ``Interferometry and Synthesis in
  Radio Astronomy'', J.W. Wiley \& Sons, Inc., 2nd Edition, 2001.
\bibitem[Thompson {\sl et al.}(1980)]{tho80} Thompson, A.R., Clark,
 B.G., Wade, C.M., and Napier, P.J. {\it Astrophys. J. Suppl.}, 44,
 151--167, 1980.
\bibitem[Napier {\sl et al.}(1983)]{nap83} Napier, P.J., and Ekers,
  R.D. {\it Proc. IEEE}, 71, 1295--1320, 1983.
\bibitem[Carlson and Dewdney (2000)]{car00} Carlson, B.R., and
  Dewdney, P.E. {\it Electron. Lett.}, 36, no. 11, 987, 2000.
\bibitem[Perley {\sl et al.}(2009)]{per09} Perley, R., Napier, P.,
  Jackson, J., Butler, B., Carlson, B., Fort, D., Dewdney, P., Clark,
  B., Hayward, R., Durand, S., Revnell, M., and McKinnon, M. {\it
  Proc. IEEE}, 97, \#8, 1448 -- 1462, 2009.
\bibitem[National Research Council (2001)]{astro2001} ``Astronomy and
  Astrophysics in the New Milennium'', National Research Council, 2001.
\bibitem[Greisen (2003)] {grei03} Greisen, E.W., ``AIPS, the VLA,
  and the VLBA'', in ``Information Handling in Astronomy -- Historical
  Vistas'', ed. A Heck, Kluwer, 109, 2003.
\bibitem[McMullin {\sl et al.}(2007)] {McMul07} McMullin, J.P., Waters, B.,
  Schiebel, D., Young, W., Golap, K., ``CASA Architecture and
  Applications'', ADASS XVI, ASP Conference Series, Ed. R.A. Shaw,
  F. Hill, \& D.J. Bell, Vol. 376, 127 -- 130, 2007.
\end{thebibliography}
\end{document}